\documentclass[useAMS,usenatbib]{mn2e}
\usepackage{epsfig}
\title[Stars in Bulges \& Ellipticals]
{On the Origin of Stars in Bulges and Elliptical Galaxies}
\author[Khochfar \& Silk]
  {S.~Khochfar,$^{1}$\thanks{sadeghk@astro.ox.ac.uk}
  J.~Silk,$^{1}$ \\
  $^1$ Department of Physics, Denys Wilkinson Building, Keble Road, Oxford OX1 3RH, United 
Kingdom\\
   }
\date{Released 2005 Xxxxx XX}

\pagerange{\pageref{firstpage}--\pageref{lastpage}} \pubyear{2005}

\def\LaTeX{L\kern-.36em\raise.3ex\hbox{a}\kern-.15em
    T\kern-.1667em\lower.7ex\hbox{E}\kern-.125emX}

\begin{document}

\label{firstpage}

\maketitle

\begin{abstract}
We investigate the stellar 
composition of bulges and elliptical galaxies as predicted 
by the CDM paradigm using semi-analytical modelling. We argue that 
spheroid stars are built up of two
main components, {\it merger} and {\it quiescent}, 
according to the origin of the stars. The merger component is formed during 
major mergers by gas driven to the centre, while the quiescent component is 
formed in gaseous discs and added later to the spheroid during major mergers.
Galaxies more massive than $M_C=3 \times 10^{10}$ M$_{\odot}$ 
\linebreak 
have on average only a
$15 \%$ merger component in their spheroids, while smaller galaxies can have 
up to $30 \%$. The merger component increases with redshift due to mergers
involving more gas. However we do not find mergers with gas fraction above
$ \sim 40 \%$ of the remnants mass. Generally the gas fraction is a 
decreasing function of the redshift at which the merger occurs and the 
mass of the remnant,
with more massive remnants having smaller gas fraction and hence smaller merger
components. This trend is independent of the environment of the galaxy with 
the only impact of the environment being that galaxies less massive than $M_C$
have slightly larger merger components in dense environments.
 The fraction of stars in bulges for galaxies 
more massive than $M_C$ is larger than $50 \%$. We find that the majority 
of stars in galaxies more massive than  $M_C$ reside within bulges and 
ellipticals independent of  redshift and that the fraction increases with 
redshift. The most massive galaxies at each redshift are elliptical galaxies.

\end{abstract}


\section{Introduction}
Bright elliptical galaxies are old stellar systems which 
formed the bulk of their 
stars at high redshift within a short time scale \citep[e.g.][]{c04,t05}. 
However, they are not a coeval family of objects as they for example
 divide into
luminous slow rotating pressure supported ellipticals, called boxy, 
and fast rotating less luminous ellipticals \citep[e.g.][]{bbf}, called disky.
Another devision of  elliptical galaxies suggests that they can be divided 
into core and power-law galaxies according to the steepness of their inner 
surface brightness profile  
\citep[e.g.][]{L95}. Furthermore, 2-dimensional integral-field spectographs
like SAURON or OASIS, reveal kinematically decoupled cores in many elliptical
galaxies \citep[e.g.][]{mc05}.

There are two competing scenarios for the formation of elliptical galaxies.
One assumes that they are formed at high redshift during a highly efficient 
starburst followed by passive evolution of the stellar component 
\citep[e.g.][]{l74,cc02}, and the other assumes the formation of an elliptical
galaxy in the merger of two galaxies \citep[e.g.][]{tt72}. In this paper, we
will test the predictions made by the latter model.

Since the early proposal of  \citet{tt72} that early type galaxies form as the
result of a binary merger between two spiral galaxies, many numerical 
simulations have been performed and compared to observations 
\citep[e.g.][and references therein]{ba92,bn03}. The general consensus is that 
it is possible to generate remnants which agree in their kinematic properties 
with observed elliptical galaxies. Recent merger simulations  use 
self-consistent cosmological initial condition \citep{kb04}, 
in simulations with only stars \citep[e.g.][]{nb03} and in simulations 
with gas, star formation and feedback \citep[e.g.][]{c05}.  Usually these 
simulations neglect   the past history of a possible progenitor 
galaxy, with the   simulations starting from idealised initial 
conditions. 

Observationally, there is evidence that the initial conditions of merging
galaxies cover a wide range, from being 'dry' \citep{vd05,be05} to being
ultra-luminous \citep[for a review see][]{sch05}. It is therefore expected 
that the stars ending up in a spheroid could not all have been formed in the
same kind of event, but must have  formed in different events which depend
 on the past history of the merging galaxies. Since merger simulations indicate
that violent relaxation is not complete and that the stars manage to remember
their initial energy and orbital angular momentum \citep{ba98},  
it is necessary to take into account the past history of a galaxy.
The main task now is to connect these past histories with the properties of
merger remnants.

Large observational studies that  collect data from several thousands 
of galaxies,
like the Sloan Digital Sky Survey,  reveal that the galaxy 
population follows remarkable trends: e.g. the surface mass density 
of galaxies increases with mass until a characteristic mass scale of 
$M_C=3 \times 10^{10}$ M$_{\odot}$ at which it becomes constant 
\citep{k03}. The constant surface mass density is mainly 
associated with elliptical galaxies and galaxies having significant bulges.
Furthermore, detailed studies of the size distribution of elliptical galaxies 
reveal that the scatter in sizes of elliptical galaxies of a given mass is 
log-normal distributed with a scatter which decreases for larger galaxy masses
\citep{sh03}. Those authors could explain the evolution of the sizes by 
assuming continued mergers of galaxies which had initially all the same mass
but sizes that followed an appropriate distribution. The origin of this 
distribution however remains unsolved. 
Another interesting observed correlation is that the surface mass density of 
galaxies in different environments only differs at low masses \citep{k04}.
 
In this paper, we address the question of where stars that end up in spheroids 
were formed. The paper is structured as follows: we begin by 
explaining the model ingredients we use, 
followed by a section on the past merging history of galaxies. Then we 
introduce our definition of the two stellar components found in spheroids
and discuss how these components evolve in the simulated galaxy population. 
In section five we present our conclusions.

\section{Galaxy formation modelling}
The main strategy behind the modelling approach we follow is first to calculate
 the collapse and merging history of individual dark matter halos, 
which is governed purely by  gravitational interactions, and secondly 
to calculate the more complex physics of the baryons inside 
these dark matter halos, including  e.g. radiative cooling of the gas, star formation, 
and feedback from supernovae by simplified prescriptions on top of 
the dark matter evolution. Each of the dark matter halos will consist of three 
main components which are distributed among individual galaxies inside  
them: a stellar, cold, and hot gas component, where the latter  is
only attributed to {\it central} galaxies, which are the most massive galaxies
inside individual halos.
In the following sections, we will describe briefly the recipes used to
 calculate these different components which 
are mainly based on recipes presented in e.g. \citet{c94,c00}, \citet{k99} 
(hereafter, K99) and 
\citet{spr01} (hereafter, S01), and we refer readers for more details 
on model implementations to their work and references therein.

Throughout this paper we use the following set of cosmological parameters:
$\Omega_0=0.3$, $\Omega_{\Lambda}=0.7$, $\Omega_b/\Omega_0=0.15$, 
$\sigma_8=0.9$ and $h=0.65$.

\subsection{Dark Matter Evolution}
We calculate the merging history of dark matter halos according to 
the prescription presented in \citet{som99a}. This approach has 
been shown to produce merging histories and progenitor distributions in 
reasonable agreement with results from N-body simulations of cold dark matter 
structure formation in a cosmological context \citep{som00}. 
The merging history of dark matter halos is reconstructed by breaking 
 each halo up into progenitors above a limiting minimum progenitor 
mass $M_{min}$. This mass cut needs 
to be chosen carefully as it ensures that the right galaxy population and 
merging histories are produced within the model. Progenitor halos
 with masses below $M_{min}$ are declared as {\it accretion} events and 
their histories are not followed further back in time. 
Progenitors labelled as accretion events should ideally not host any 
significant galaxies in them and be composed mainly of primordial hot gas 
at the progenitor halo's  virial temperature. 
The mass scale at which this is the case can in principle be 
estimated from the prescriptions of supernova feedback and reionization 
presented in section \ref{cool}.
However,  to achieve a 
good compromise between accuracy and computational time, we instead estimated 
$M_{min}$  by running several simulations with different resolutions 
and chose the resolution for which results in the galaxy mass range of 
interest are independent of the specific choice of $M_{min}$. 
Changing the mass resolution mainly affects our results at low galaxy mass 
scales as shown in Fig.1, leaving massive galaxies nearly unaffected.
Throughout this paper we will use $M_{min}=2 \times 10^{9}$ M$_{\odot}$  
which produces numerically stable results for galaxies with stellar masses 
$M_{*} \geq 10^9$ M$_{\odot}$.
\begin{figure}
  \begin{center}
    \includegraphics[width=8cm,angle=0]{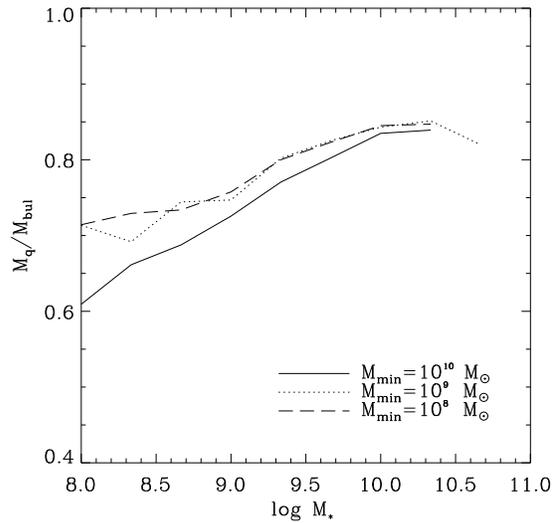} 
    \medskip
    \caption{Average mass fraction of disc stars populating spheroids 
      depending on the mass resolution $M_{min}$ of the merger tree. 
      \label{fig1}}  
  \end{center}
\end{figure}

\subsection{Baryonic Physics}
As mentioned above, once the merging history of the dark matter 
component has been calculated, it is
 possible to follow the evolution of the baryonic content in these halos 
forward in time. We assume each halo  consists of three components: 
hot gas, cold gas and stars, where the latter two components can be distributed 
 among individual galaxies inside a single  dark matter halo. The stellar components 
of each galaxy are additional divided into bulge and disc, to allow 
morphological classifications of model galaxies. In the following, we 
describe how the evolution of each component is calculated. 

\subsubsection{Gas Cooling \& Reionization}\label{cool}
Each branch of the merger tree starts at a progenitor mass of $M_{min}$ and 
ends at a redshift of $z=0$. Initially, each halo is occupied by hot 
primordial gas which was captured in the potential well of the halo and shock 
heated to its virial temperature 
$T_{vir}=35.9\left[V_c/(\mbox{km s}^{-1}) \right]^2$ K, where $V_c$ is the 
circular velocity of the halo \citep[K99]{wf91}. Subsequently this hot 
gas component is allowed to radiatively cool and settles down into a
rotationally supported gas disc at the centre of the halo, which we identify 
as the central galaxy \citep[e.g][]{s77,wr78,wf91}.  
The rate at which hot gas cools down is estimated by calculating the 
cooling radius inside the halo using the cooling functions provided by 
\citet{sd93} and the prescription in S01. In the case of a merger 
between halos we assume that all of the hot gas present in the progenitors 
gets shock heated to the virial temperature of the remnant halo, and that gas 
can only cool down onto the new central galaxy which is the central galaxy 
of the most massive progenitor halo. The central galaxy of the less massive 
halo will become a satellite galaxy orbiting inside the remnant halo. In this 
way, a halo can host multiple satellite galaxies, depending on the 
merging history of the halo, but will always only host one central galaxy onto 
which gas can cool. The cold gas content in  satellite galaxies  is given by 
the amount present when they first became satellite galaxies and does 
not increase, instead it decreases due to ongoing star formation and supernova 
feedback.

In the simplified picture adopted above, the amount of gas available to 
cool down is only limited by the universal baryon fraction 
$\Omega_b h^2=0.024$ \citep{sper03}. However, in the presence of a 
photoionising background the fraction of baryons captured in halos is reduced
\citep[e.g.][]{ns97,g00,ben02} and we use the recipe of \citet{som02}, 
which is based on a fitting formulae derived from hydrodynamical simulations 
by \citet{g00}, to estimate the amount of baryons in each halo. For the epoch 
of  reionisation, we assume $z_{reion}=20$, which is in agreement with 
observations of  
the temperature-polarisation correlation of the cosmic microwave background 
by \citet{ko03}. 

\subsubsection{Star formation in Discs and Supernova Feedback}\label{sf}
Once cooled gas has settled down in a disc, we allow for fragmentation and 
subsequent star formation according to a parameterised global 
Schmidt-Kennicutt law  \citep{ken98} of the form 
$ \dot{M}_{*}=\alpha M_{cold}/t_{dyn,gal}$, where $\alpha$ is a free parameter
describing the efficiency of the conversion of cold gas into stars, and 
$t_{dyn,gal}$ is assumed to be the dynamical time of the galaxy and is 
approximated to be 0.1 times the dynamical time of the dark matter halo (K99).
As in K99 we allow star formation only in halos of $V_c < 350$ km/s to avoid 
too bright central galaxies in clusters. 

Feedback from supernovae plays an important role in regulating star 
formation in small mass halos and in preventing too massive satellite 
galaxies from forming. We implement feedback based on the prescription presented in K99 with
\begin{equation}
  \Delta M_{reheat}=\frac{4}{3} \epsilon \frac{\eta_{SN} E_{SN}}{V_{c}^{2}} 
  \Delta M_*. 
\end{equation}
Here we introduce a second free parameter $\epsilon$ which represents our 
lack of knowledge on  the efficiency with which the energy from supernovae is 
going to reheat the cold gas.  The expected number of supernovae 
per solar mass of stars formed is given by $\eta_{SN}=5 \times 10^{-3}$,
taken as the value for the Scalo initial mass function \citep{sca86},  
 and $E_{SN}=10^{51}$ erg  is the energy output from each supernova. We take 
$V_c$ as the circular velocity of the halo in which the galaxy was  
last present as  a central galaxy.

\subsubsection{Galaxy Mergers}
We allow for mergers between galaxies residing in a single halo. As mentioned 
earlier, each halo is occupied by one central galaxy and a number of 
satellite galaxies depending on the past merging history of the halo. 
Whenever two halos merge, the galaxies inside  them are going to merge on a 
time-scale which we calculate by estimating the time it would take the 
satellite to reach the centre of the halo under the effects of 
dynamical friction. Satellites are assumed to merge only with 
central galaxies and we set up their orbits in the halo according to the 
prescription of K99, modified to use the Coulomb logarithm 
approximation of S01.

If the mass ratio between the two merging galaxies is $M_{gal,1}/M_{gal,2} 
\leq 3.5$ ($M_{gal,1} \geq M_{gal,2}$) we declare the event as a 
{\it major} merger and the remnant will be an elliptical galaxy and the 
stellar components and the gas will be treated according to the prescriptions 
below. In the case of {\it minor} merger $M_{gal,1}/M_{gal,2} > 3.5$ 
 the cold gas in the disc of the smaller 
progenitor is assumed to settle down in the gas disc of the remnant and its 
stars contribute to the bulge component of the remnant (e.g. K99).

\subsubsection{Formation of Ellipticals and Bulges}
\citet{tt72} suggested that major mergers will lead 
to the formation of elliptical galaxies. Indeed 
detailed numerical simulations in the 
last decade seem to support this hypothesis \citep[e.g.][and reference therein]
{ba92,bn03}, and we will assume in the following that major 
mergers disrupt the 
discs in progenitor galaxies, as seen in various numerical simulations,  
 and relax to a spheroidal 
distribution. During the merger, any cold gas in the discs of the progenitor 
galaxies is assumed to be funnelled into the centre of the remnant where it
ignites a starburst which transforms all of the cold gas into stars 
contributing to the 
spheroidal component (e.g. K99; S01; and references therein). The second 
assumption is certainly a 
simplification of what might happen since we neglect the possibility  that not all of the 
cold gas is funnelled to the centre but some fraction of it can e.g. 
settle down into an extended disk 
 which continues growing inside-out by fresh supply of gas from tidal 
tails \citep[e.g.][]{bh91,mih96,ba01,b02}. 
The results of \citet{b02} indicate that $40\% - 80\%$ 
of the initial gas in the disc could end up in the central region of 
the remnant and be consumed 
in a starburst. The exact amount is somewhat dependent on the merger 
geometry and on the mass ratio of the merger. Unfortunately, a large survey  
investigating the gas inflow to the centres of merger remnants is 
not available to date so that we use the simplified approach of assuming that
all cold gas gets used up in the central starburst.
This prescription for the fate of 
the cold gas  results in an overestimate of the spheroid masses and 
an underestimate 
of the secondary disc components in our model. This  is not very significant  
for massive ellipticals since they are mainly formed in relatively weak 
dissipative mergers \citep{kb03}. 
\begin{figure}
  \begin{center}
    \includegraphics[width=8cm,angle=0]{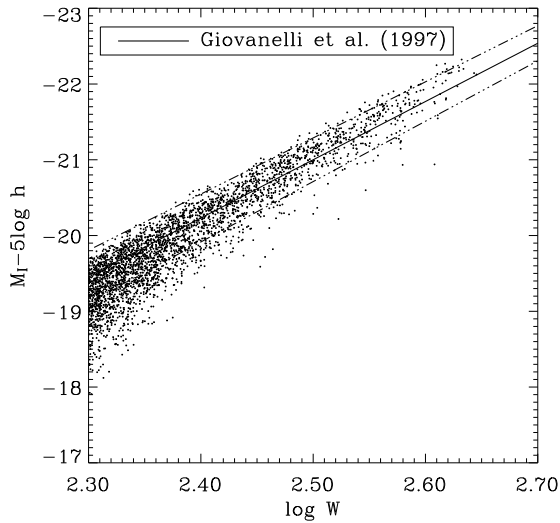} 
    \medskip
    \caption{ Predicted $I$-band Tully-Fisher relation for Sb/Sc galaxies 
      in our simulation. The solid line is the fit to the observed data by 
      \citet{gio97} and dashed lines give their estimate for the scatter.}
      \label{mod3}  
  \end{center}
\end{figure}

Another simplifying assumption is that we neglect the 
 feeding of super massive black holes in the centre of the remnant or feedback 
effects on the gas from the central source. However, \citet{hk00} estimate 
that a 
cold gas mass fraction of less than 1\% accreted onto the black hole is 
sufficient 
to recover the $M_{\bullet}-\sigma$ relation and we therefore neglect this 
effect on the amount of gas available for the central starburst. A larger 
effect on our results arises from  the fact that we neglect feedback  
from the central source into the surrounding intergalactic medium. As a 
consequence, our estimates of stars formed in a central starburst will be too 
high and should be viewed as upper limits.

\begin{figure}
  \begin{center}
    \includegraphics[width=8cm,angle=0]{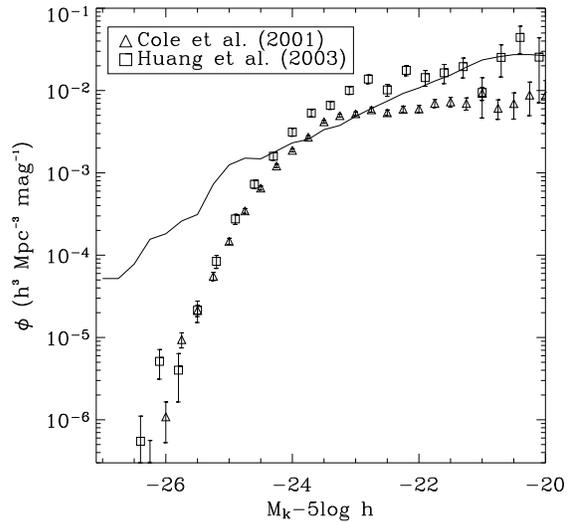} 
    \medskip
    \caption{ The predicted $K$-band luminosity function
      in our simulation in comparison to the observations by \citet{col01}
      and \citet{hua03}.}
      \label{mod2}  
  \end{center}
\end{figure}

\subsubsection{Free parameters and predicted present-day properties of modelled galaxies}
We normalise our two model parameters for the star formation 
efficiency $\alpha$ 
and feedback efficiency $\epsilon$  
by matching the $I-$band Tully-Fisher relation 
of \citet{gio97} and requiring that spiral central galaxies of 
halos with circular velocity
$V_C=220$ km/s have on average $10^{11}$ M$_{\odot}$ of stars and 
a few times $10^9$ M$_{\odot}$ of cold gas \citep{som99b}. The best fit model 
parameters we get in this way are $\epsilon=0.2$ and $\alpha=0.1$.

Figure \ref{mod3} shows our predicted $I$-band Tully-Fisher relation for 
Sb/Sc galaxies in comparison to the observed one by \citet{gio97}. We selected 
Sb/Sc galaxies following \citet{sd86} according to their bulge to total 
light in the range $1.5 \leq M_{B,bulge}- M_{B,total} \leq 2.2$.

\begin{figure}
  \begin{center}
    \includegraphics[width=8cm,angle=0]{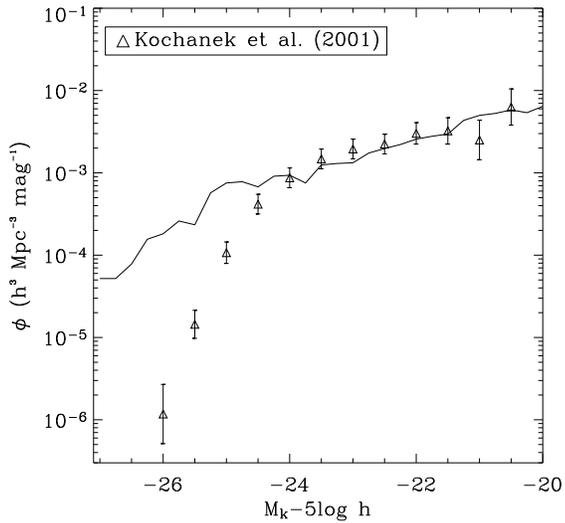} 
    \medskip
    \caption{ The predicted $K$-band luminosity function
      in our simulation for early-type galaxies in comparison to the 
      observations by \citet{ko01}.}
      \label{mod1}  
  \end{center}
\end{figure}

The model predictions for the $K$-band luminosities of the overall 
galaxy population and the early-type galaxy luminosity are shown in Fig. \ref{mod1} \& ref{mod2}, respectively. We choose early-type galaxies to have  bulge 
to total light ratios with $1 \leq M_{B,bulge}- M_{B,total}$, 
which includes S0 galaxies according to \citet{sd86}. Our luminosity 
functions somehow over predicts the abundance of very luminous objects, which is
a known problem and previously discussed in \citet{ben03}. The solution to  
this problem is commonly believed to come from AGN-feedback. \citet{cro06} 
who use semi-analytic modelling based on the same recipes that we do find a 
luminosity function very similar to ours (see Fig.8 of their paper). However,
 the inclusion of AGN-feedback in their models reduces the amount of very 
luminous galaxies and fits the data very well at the high mass end. As for 
our results presented in this paper the neglect of AGN-feedback results 
in a few to bright and too massive early-type galaxies as can be seen in Fig. 
\ref{mod2}. However, these galaxies are rare and extreme cases not considered 
in our analyses. Our early-type luminosity fiction fits the data very well 
over most of the considered range and we therefore delay further discussion 
of AGN-feedback to a forthcoming paper (Khochfar \& Silk in prep.).

\section{Past merging history of progenitor galaxies}
\begin{figure}
  \begin{center}
    \includegraphics[width=8cm,angle=0]{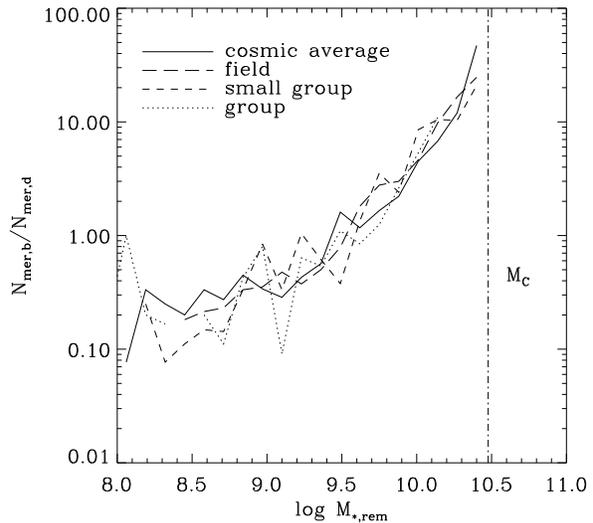} 
    \medskip
    \caption{Ratio of the number of mergers between galaxies with bulges 
      and without bulges as a function of the remnant mass found in our 
      simulations. The solid line shows the results for the cosmological 
      mass function of \citet{jen01}, and the other lines examine results 
      for different 
      environments. The vertical dot-dashed 
      line marks the characteristic mass scale
      $M_C=3 \times 10^{10}$ M$_{\odot}$ reported in \citet{k03}
      \label{fig2}}  
  \end{center}
\end{figure}

The composition of stars in bulges and elliptical galaxies is very 
dependent on the previous merging history of the progenitors and their 
ability to cool gas and regrow large stellar discs in between major mergers.
\citet{kb03} showed that based on the stellar mass of the remnant, 
a transition 
 appears at which regrowing of a stellar disc becomes  inefficient. This
leads to major mergers  mainly between bulge-dominated galaxies with too low 
fractions of available cold gas to ignite central starbursts, 
sometimes called {\it "dry" mergers}. It appears that such a transition is 
necessary to explain  properties like e.g. the isophotal shape of massive 
elliptical galaxies \citep{kb05}. 
To stress the importance of taking into account the past merging history 
of progenitors,  
we investigate in Fig. \ref{fig2} the question of  the relative number of 
mergers between galaxies without bulges to the number of mergers between 
galaxies with bulges, 
depending on the masses of the remnants. As can be seen, it 
becomes very unlikely to find mergers between two pure disc galaxies that 
lead to very massive remnants. At masses above 
$M_* \simeq 3 \times 10^{9}$ M$_{\odot},$
the majority of mergers take place between galaxies that have already experienced 
a major merger in their past. It is interesting to note that we do not
find any mergers between bulge-less disc galaxies in our simulation 
at masses larger than the characteristic mass scale $M_C$, suggesting
 that the effect of constant surface mass density $\mu_*$ in elliptical 
galaxies more massive than $M_C$ is closely related to the bulge fractions 
in the progenitor galaxies. Inspecting the environmental dependence, we find 
that the ratios evolve independently of the environment of the galaxy
and that bulge-less mergers stop  occurring at lower masses in dense 
environments like groups and clusters. This is a consequence of the 
higher merger fractions at earlier times in these environments \citep{kb01}.

\section{Merger and Quiescent Components in Spheroids}
The populations of stars present in a spheroid at a given time in its evolution
are a composite of stars formed in various progenitors at different times and 
in different modes. Here we distinguish between two modes, the merger-induced
{\it 'merger component'} and the disc mode {\it 'quiescent component'}. The
merger component is a composite of all stars that were formed because of the
consumption of cold gas during major mergers in progenitor galaxies. On the 
other hand, the quiescent component is the amount of stars formed in gaseous 
discs according to the Schmidt-Kennicutt law during the evolution of 
the progenitor galaxies. When calculating each 
component for a spheroid, we sum up the mass of the merger component so far and
subtract it from the spheroid mass to get the quiescent component. 

To investigate systematic trends in the fraction of merger-induced
 and quiescent stellar components in elliptical galaxies and bulges, we  
use in the following the conditional distribution 
\begin{equation}
  p(M_{q}/M_{bul}|M_*)=\frac{\phi(M_{q}/M_{bul},M_*)}
  {\int \phi(M'_{q}/M'_{bul},M_*) d(M'_{q}/M'_{bul})},
\end{equation}
where $M_q$ is the stellar mass in the spheroid previously formed quiescently, 
$M_{bul}$ the mass of the spheroid and $M_*$ the total 
stellar mass of the galaxy. The merger component in the spheroid is then 
simply given by $1-M_{q}/M_{bul}$.
If not stated otherwise, this distribution is derived for 
the simulated galaxy population at redshifts $z\leq 0.3$.  
\begin{figure}
  \begin{center}
    \includegraphics[width=8cm,angle=0]{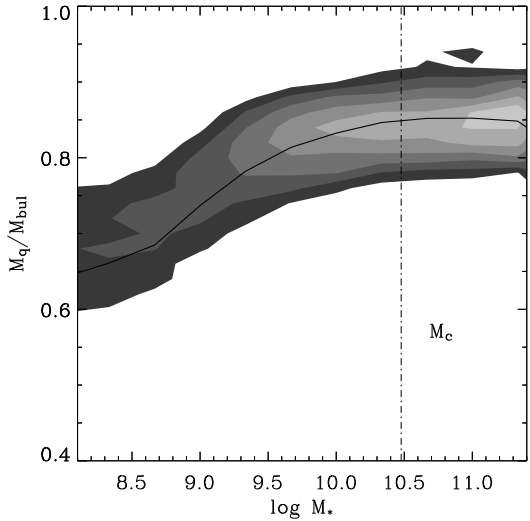} 
    \medskip
    \caption{Conditional distribution of the fraction of quiescent stars 
      ending up in spheroids as a function 
      of the total galaxy mass for the overall galaxy population at redshift 
      $z \leq 0.3$. The solid line shows the mean of the distribution and the
      vertical dot-dashed line marks the characteristic mass scale $M_C$.
      \label{fig3}}  
  \end{center}
\end{figure}

The quiescent component of stars in spheroids is getting more dominant 
with galaxy mass up to masses around $M_C$ when it starts 
becoming a constant fraction of the stars in the spheroid (Fig. \ref{fig3}).  
The behaviour at galaxy masses below $M_C$ can be understood by acknowledging
that massive spheroids form late in the hierarchical galaxy formation paradigm,
thereby allowing more stars to be formed in progenitor discs between individual
major merger events, and that feedback from supernovae is less efficient in 
reheating cold gas in discs which are embedded in massive halos. It is worth 
noting that even spheroids in galaxies as small as $M_*=10^9$ M$_{\odot}$ 
have on average only $\sim 30 \%$ of their stars being formed in past 
merger-triggered central 
starbursts, indicating that the main mode of star formation is taking place 
in discs. The quiescent fraction of stars can get as large as $\sim 85 \%$
at masses above $M_C$, where it then becomes constant.
\begin{figure}
  \begin{center}
    \includegraphics[width=8cm,angle=0]{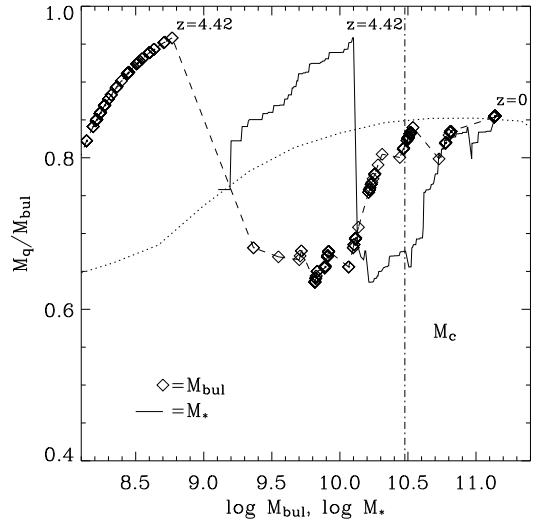} 
    \medskip
    \caption{Trajectory of a representative galaxy with final mass 
      $M_*=1.5 \times 10^{11}$ M$_{\odot}$ at $z=0$. We show the evolution as 
      a function of the bulge mass $M_{bul}$ and as
      a function of the total stellar mass $M_*$,  indicated by the symbols
      and the solid line, respectively. The average of the overall galaxy 
      distribution is displayed by the dotted line, and the 
      vertical dot-dashed  line 
      marks the characteristic mass scale $M_C$. Please note that the 
      trajectory as a function of total stellar mass starts around 
      $M_* \sim 10^9$ M$_{\odot}$ because we start following 
      trajectories after the first occurrence of a major merger between
      progenitor galaxies. For guidance we label the bulge and total 
      stellar mass at the first occurrence of a
      significant central star burst with the 
      corresponding redshift of that event $(z=4.42)$ and at a redshift 
      of $z=0$. 
      \label{fig4}}  
  \end{center}
\end{figure}

\subsection{Redshift Evolution}\label{redsh}
The distribution found for the galaxy population at $z \leq 0.3$ is likely
to be dependent on redshift. Galaxies at high redshifts tend to have a higher 
fraction of cold gas available in their discs than their counterparts 
at low redshift when they participate in mergers  as shown in Fig. \ref{mod5}. 
This is mainly due to the 
time available to make stars in the disc before the major merger and the very 
short cooling times for the hot gas.  To 
investigate the build up of the distribution at low redshift, we look at 
 the trajectory of a galaxy  defined by its positions 
in the $M_q/M_{bul}-M_*$ plane through time. 
 Since the present day value of $M_q$ is integrated over the entire 
history of a galaxy, we calculate the trajectories by 
summing up at each redshift the
quiescent stellar components in the individual progenitor bulges and 
ellipticals and divide it by the sum of the 
stellar mass of the progenitor spheroids:
\begin{equation}
  \frac{M_q(z)}{M_{bul}(z)}=\frac{\sum_{i=1,N_{prog}}M_{q,i}(z)}
       {\sum_{i=1,N_{prog}}M_{bul,i}(z)}.
\end{equation}
 Note that elliptical 
galaxies might contain a small disc which is in the process of regrowing, and 
that we do not attribute these stars to the spheroid when calculating the 
trajectory.
A typical trajectory for an elliptical galaxy more massive than $M_C$ 
is presented in Fig. \ref{fig4}. This galaxy experienced a
 gas-rich major merger
at a redshift of $z=4.42$ between two progenitors leading to a significant 
decrease in the quiescent fraction $M_q/M_{bul}$. 
The significant decrease is not only due to the 
amount of cold gas available in the merger but also to the fact that most of 
the stars in the individual progenitors are still in their discs, and that 
only a few, if any, galaxies consist of spheroids, which makes a single major 
merger have a large impact. Once significant spheroids in the population
of progenitors start  to 
exist, major mergers between two progenitors tend to 
lower the overall fraction of quiescent stellar components only slightly. 

The main mode by which $M_q/M_{bul}$ increases 
is by minor merging with satellite galaxies. These satellite galaxies, though 
smaller in mass than their central galaxies, have grown stellar discs
which are massive enough to increase  $M_q/M_{bul}$. That minor mergers are 
important becomes evident by noting that they are around an order of 
magnitude more frequent than major mergers. Once the bulge mass reaches roughly
$M_C$ the ratio $M_q/M_{bul}$ is around the average value inferred from the 
overall distribution at that redshift. Subsequently the mergers, minor as well
 as major,  only create a small scatter around this value, consistent 
with the assumption of bulge-dominated mergers.
\begin{figure}
  \begin{center}
    \includegraphics[width=8cm,angle=0]{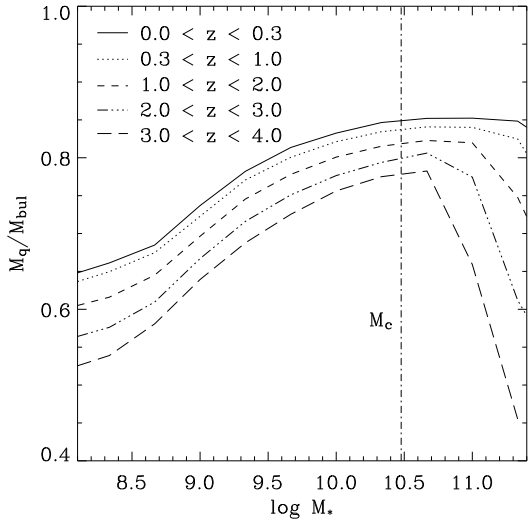} 
    \medskip
    \caption{The evolution of the mean value $M_q/M_bul$ of the 
      conditional distribution 
      $ {p}(M_{q}/M_{bul}|M_*)$ as a function of stellar mass 
      $M_*$  in five different redshift bins between $z=0$ and $z=4$.
      The vertical dot-dashed line marks the characteristic mass scale $M_C$.  
      \label{fig5}}  
  \end{center}
\end{figure}

The trajectory in Fig. \ref{fig4} suggests that the mean value of the 
conditional distribution will change with time, in co-evolution with 
the merger 
activity. Fig. \ref{fig5} compares the mean value of $M_{q}/M_{bul}$ 
found in the conditional distributions $ {p}(M_{q}/M_{bul}|M_*) $ 
in five different redshift bins. 
At earlier times, the means show systematic lower fractions of quiescent 
components in their spheroids. This is connected to the higher fraction of
available gas in major mergers at earlier times as can be seen in Fig. 
\ref{mod5}. 
 The different 
averages show similar behaviours with stellar mass below $M_C$ only differing 
by constant offsets and a slightly steeper increase of $M_q/M_{bul}$ 
toward $M_C$. Above $M_C$, the starburst component in spheroids starts to 
increase again by up to $\sim 30 \%$ depending on the redshift, 
a result of massive galaxies at high redshift  likely being 
formed in a very 
gas-rich merger event, or having bulges that were formed very early 
on. At this point, it is not clear if this trend at high redshifts 
is real or just a product of not including feedback from super-massive black 
holes during major mergers and we will defer discussion of
 possible effects of AGN-feedback to a following paper 
(Khochfar \& Silk in prep.). However, note that we usually do not find 
more than a few galaxies in the highest mass bins, which makes finding
 such galaxies  very unlikely.
From the different 
curves in Fig. \ref{fig5}, it is not obvious whether $M_C$ plays the same 
fundamental role at all redshifts or if the characteristic mass scale changes
with redshift. In any case, the differences according to our simulations 
are not very large for the overall galaxy population.
\begin{figure}
  \begin{center}
    \includegraphics[width=8cm,angle=0]{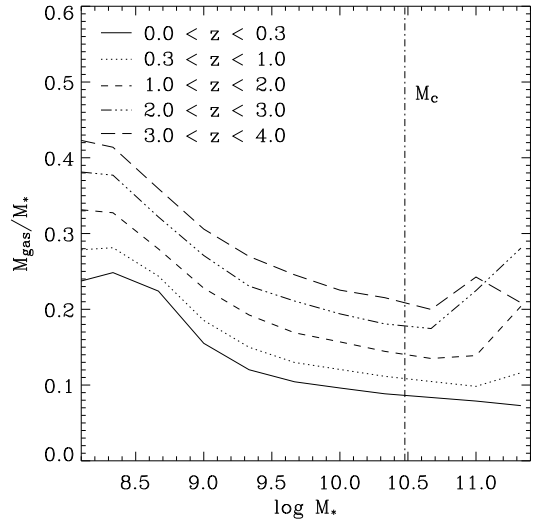} 
    \medskip
    \caption{The evolution of the mean gas fraction in progenitors of major 
      mergers as a function of stellar mass 
      $M_*$  in five different redshift bins between $z=0$ and $z=4$. 
      Please note that $M_*$ is the stellar mass of the merger remnant.
      \label{mod5}}  
  \end{center}
\end{figure}

\begin{figure}
  \begin{center}
    \includegraphics[width=8cm,angle=0]{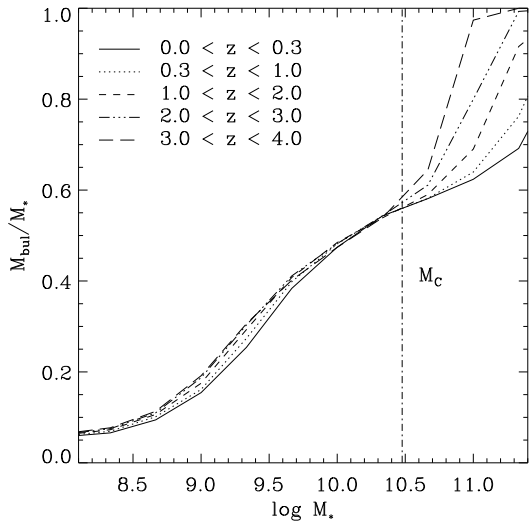} 
    \medskip
    \caption{Bulge to total stellar mass fraction 
      in five different redshift bins between $z=0$ and $z=4$. 
      We calculated the fraction by summing up all stars present in 
      spheroids at a given redshift and mass scale and dividing this value 
      by the total amount of stars available at this mass scale. 
      The vertical dot-dashed line marks the 
      characteristic mass scale $M_C$.
      \label{fig5b}}  
  \end{center}
\end{figure}

To understand the shape of the curves in Fig. \ref{fig5} 
it is important to understand the extent to which
 the evolution in $M_q/M_{bul}$ 
at each redshift and mass scale is driven by galaxies with different 
morphologies. In Fig. \ref{fig5b} we compare the overall stellar mass in 
spheroids divided by the total stellar mass present at different redshifts 
and masses. The closer to $M_{bul}/M_* =1$, the more likely it is that most of 
the galaxies at a given mass scale are elliptical galaxies which have recently 
formed and did not regrow significantly large discs since then. 
At each redshift, our 
model predicts that the most massive galaxies, which are less massive with 
increasing redshift, are  recently formed elliptical galaxies. 
As a consequence, the quiescent component in these galaxies resembles the 
average composition that results from  major mergers at this redshift. 
Going to smaller galaxies, the bulge mass 
fractions decrease, indicating that more mass is found in stellar discs, 
which in turn means that the bulges must have formed earlier to allow
for the growth of large stellar discs. As  argued above, early mergers are 
more gas-rich and therefore lead to smaller ratios of $M_q/M_{bul}$. This, and 
the fact that galaxies with increasing mass have a larger number of 
minor mergers, which in general increase the quiescent component of bulges,
accounts for the increase in the quiescent component in spheroids with 
increasing mass.

It is interesting to note that Fig. \ref{fig5b} suggests that the 
characteristic mass scale $M_C$ approximately  marks the 
transition point at which most of the stars start to be in spheroids 
independent of the redshift. Below $M_C$, the average distribution of stars
in bulges and discs is roughly independent of the redshift, and shows the same
mass dependence. Only at galaxy masses 
above $M_C$ does  the distribution of stars in spheroids and discs become 
significantly redshift-dependent. This behaviour can be attributed to two 
features of the CDM paradigm: first, the 
decline in the merger rate going to lower redshifts and the connected ability
of stellar discs to grow longer without being disrupted, and second,  most 
massive galaxies at a given redshift are likely 
to be those which formed in a major merger, resulting in an elliptical galaxy. 

To summarise, the build up of bulges and elliptical galaxies in a merger-driven
model suggests that early gas-rich mergers lead to progenitors with 
a high fraction of stars in their spheroids, and which were formed  during a
dissipative central starburst event. This fraction is significantly 
higher than that found at later times in spheroids. While the progenitors
 of a present-day galaxy continue to grow stellar discs, minor mergers occur. 
These events tend to increase the quiescent fraction of stars in spheroids by
contributing their disc stars to the spheroid of the remnant. These minor 
mergers occur mainly during the peak times in the merger activity which is 
dependent on the environment that the galaxy resides in \citep{kb01}. 
As soon as the bulge mass of the progenitor galaxies reaches 
$ \sim M_C$, only small changes occur. However, depending on the morphology 
of the resultant galaxy the final stages might differ as well as the 
characteristic mass scale (Khochfar \&Silk, in preparation). 
 
\subsection{Environmental Dependence}
In paragraph \ref{redsh} we presented the evolution 
of each individual stellar component in spheroids. This evolution is very 
dependent on the merger activity that galaxies experienced in their past. 
It is to be expected that the different evolutions of the 
merger rates with redshift in field and high density environments leave 
a characteristic signature in the merger component of spheroids.

\begin{figure}
  \begin{center}
    \includegraphics[width=8cm,angle=0]{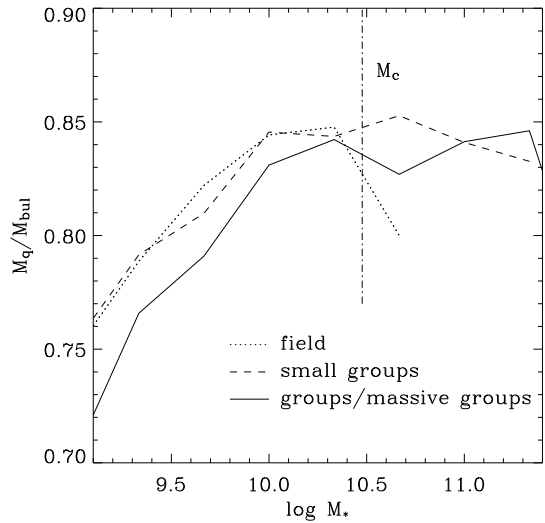} 
    \medskip
    \caption{The evolution of the mean value $M_q/M_bul$ of the
      conditional distribution 
      ${p}(M_{q}/M_{bul}|M_*)$ as a function of stellar mass 
      $M_*$  in three different environments. We defined environments by the
      the mass of the dark halo the galaxies reside in. Halos with 
      $10^{12}$ M$_{\odot}$, $10^{13}$ M$_{\odot}$ and 
      $10^{14}$ M$_{\odot}$ are classified as field, small group and 
      group/massive group environments, respectively.
      \label{fig8}}  
  \end{center}
\end{figure}

 In Fig. \ref{fig8}, we compare the fraction of quiescent components
in the spheroids of galaxies found in different environments in our simulation.
We classify environments into three different categories according to the 
mass of the dark matter halo the galaxies reside in. Galaxies in dark matter 
halos of $10^{12}$ M$_{\odot}$, $10^{13}$ M$_{\odot}$ and 
$10^{14}$ M$_{\odot}$ are classified as field, small group, and 
group/massive group environments in our simulation. The average fraction of 
quiescent components in the bulges of spheroids less massive than $\sim M_C$
is systematically higher in field environments meaning their merger 
components are smaller.  Furthermore we find in our 
simulations that there is no systematic difference for galaxies above 
$\sim M_C$. The difference  between environments diminishes, the closer 
the galaxy mass is to $M_C$. Recalling the results of section \ref{redsh},
it becomes clear that massive galaxies in different environments are 
unaffected because the continuous growing by mergers, minor as well as major,
  erases the memory of the first merger epoch in different
environments. Only those galaxies  significantly less massive than 
$\sim M_C$ did not grow by many  mergers after their initial mergers, but
survived until today without much interactions as satellite galaxies. 
Since the initial merger is very much environmental dependent, gas-rich early 
mergers in high density environments and less gas-rich late mergers in field 
environments, differences arises between environments.

\section{Discussion and Conclusion}
In this paper we have investigated the origin of stars ending up in bulges 
and elliptical galaxies for systematic trends with mass, redshift and time of 
the last major merger.  We  divide stars in spheroids 
into two categories, merger and quiescent, according to their formation mode.

The merger component is 
associated with major mergers in which cold gas in progenitor discs is driven
 to the centre of the remnant and a central starburst is ignited. Our 
simulations predict that the fraction of the merger component in 
spheroids is an increasing 
function of formation redshift and a decreasing function of spheroid mass. 
The latter is due to the longer time that progenitors of massive galaxies 
have to grow discs, and an
enhanced number of minor mergers, which generally decreases
the fraction. The higher fraction at earlier formation times is a result of 
the short cooling time scales for gas and the insufficient build up of stellar 
discs before the final major merger. Our simulations predict that the average
merger component in the local galaxy population can be as high as $30 \%$ 
of the bulge mass in galaxies with $M_* \sim 10^9$ M$_{\odot}$ and as low as
$15 \%$ in galaxies with $M_* > M_C=3 \times 10^{10}$ M$_{\odot}$. 

The quiescent component is associated with stars previously formed in gaseous 
discs according to the Schmidt-Kennicutt law and later being added to the 
spheroid during major mergers. We find that the fraction of quiescent stars 
dominates bulges and elliptical galaxies. For galaxies 
above the critical mass scale, the fraction of quiescent stars 
becomes constant 
at a value of $M_q/M_{bul} \sim 85 \%$, the result of mergers, 
minor as well as major, only introducing small deviations as soon as galaxies 
are more massive than $M_C$.  

The fraction of stars in bulges and discs shows a remarkable relation with the 
galaxy mass scale. At galaxy masses below $M_C$, we find only a weak redshift 
dependency and a decreasing fraction of stars in bulges with decreasing mass.
However, at masses above $M_C$ we find a strong redshift dependence. At higher
redshifts, the fraction of stars in bulges becomes larger in massive galaxies.
This means that the most massive galaxies around at each redshift are likely 
to be elliptical galaxies and that only  at later times do massive spiral 
galaxies start occurring. 

Furthermore, our
simulations indicate that mergers leading to remnants with mass larger than 
$M_C$ always include progenitors with bulges and that more than $50 \% $ 
of the mass in progenitors of mass $M_* \sim M_C$ is in bulges. This has an 
important influence on the role of AGN-feedback as these bulges will harbour 
super-massive black holes.

Looking at the environmental dependence, we find that only galaxies less 
massive than $M_C$ show differences. This is mainly related to the different 
epoch at which these galaxies assembled. Assembly is faster in high density 
environments resulting in larger merger components in bulges. Galaxies above
$M_C$ do not show any significant environmental dependency because of continued
merger activity  in contrast to smaller galaxies.

Recently \citet{sh05} suggested that mergers between extremely gas-rich discs
could lead to the formation of a disc galaxy during a major merger. The authors
 argued that these kind of mergers are likely to occur at high redshifts. 
However, our simulations indicate that the maximum gas fraction is at
 $\sim 40 \%$ in mergers at high redshift and most likely between 
$20 \% - 30 \%$
for a wide range of masses and redshifts. In general the gas fraction 
in mergers is a decreasing function of mass. At very large masses and at high 
redshifts a small population of gas rich massive mergers occurs, which 
tends to slightly deviate from this trend. A reason for these events in our 
simulation could be associated with missing feedback from AGNs. Those massive 
gas-rich mergers at high redshift are likely to have progenitors with 
black holes.

Simulations by \citet{sh05} show that in general the merger component is more
centrally concentrated than the quiescent component. The effective radius 
of the merger component in their simulation is $\sim 5.7$ times smaller than 
that of the quiescent component. This suggest an interesting behaviour for the 
sizes of two galaxies of the same mass but different merger component. The 
remnant from the merger including less gas and hence less merger component 
will have a larger sizes. As we have shown above remnants at higher redshifts 
have larger merger fractions. This and the observational indication that disc 
galaxies  at higher redshifts are smaller \citep[e.g][]{tru05} might be 
able to account for the  observed size-evolution of elliptical galaxies 
\citep{da05,tru05}.
\newline

We would like to thank Ignacio Trujillo and  Emanuele Daddi for pointing
out the high redshift data.
SK acknowledges funding by the PPARC Theoretical Cosmology Rolling Grant.



\label{lastpage}

\end{document}